# Multi-photon transitions and Rabi resonance in continuous wave EPR

Alexander P. Saiko [a], Ryhor Fedaruk [b], Siarhei A. Markevich [a]

[a] Scientific-Practical Materials Research Centre NAS of Belarus, Minsk, Belarus

[b] Institute of Physics, University of Szczecin, 70-451, Szczecin, Poland

E-mail addresses: saiko@ifttp.bas-net.by; fedaruk@wmf.univ.szczecin.pl

**Abstract.** The study of microwave and radiofrequency multi-photon transitions in continuous wave (CW) EPR spectroscopy is extended to a Rabi resonance condition, when the radio frequency of the magnetic-field modulation matches the Rabi frequency of a spin system in the microwave field. Using the non-secular perturbation theory based on the Bogoliubov averaging method, the analytical description of the response of the spin system is derived for all modulation frequency harmonics. When the modulation frequency exceeds the EPR linewidth, multi-photon transitions result in sidebands in absorption EPR spectra measured with phase-sensitive detection at any harmonic. The saturation of different-order multi-photon transitions is shown to be significantly different and to be sensitive to the Rabi resonance. The noticeable frequency shifts of sidebands are found to be the signatures of this resonance. The inversion of two-photon lines in some spectral intervals of the out-of-phase first-harmonic signal is predicted under passage through the Rabi resonance. The inversion indicates the transition from absorption to stimulated emission or vice versa, depending on the sideband. The manifestation of the primary and secondary Rabi resonance is also demonstrated in time-resolved steady-state EPR signals formed by all harmonics of the modulation frequency. Our results provide a theoretical framework for future developments in multi-photon CW EPR spectroscopy, which can be useful for samples with long spin relaxation times and extremely narrow EPR lines.

## 1. Introduction

Modulation of the static magnetic field and phase-sensitive detection is traditionally used to improve the signal-to-noise ratio in continuous wave (CW) measurements of electron paramagnetic resonance (EPR) signals. In these measurements electron spins interact with transverse microwave (MW) and longitudinal radiofrequency (RF) fields with the frequencies $\omega_{mw}$ and $\omega_{rf}$, respectively. EPR spectra obtained by this technique have been described by using modified Bloch equations [1,2]. This description was revisited and modulation effects in the CW EPR spectra were explained using the concept of multi-photon transitions [3,4]. The bichromatic field transforms a two-level spin system into a dynamical multilevel system and excites multi-photon transitions in this field-modified spin system. These transitions with frequencies $\omega_{mw} + k\omega_{rf}$ ($k = 0, \pm 1, ...$), where one photon of the MW field is absorbed simultaneously with absorption or emission of an arbitrary number $k$ of photons of the modulation RF field, have been studied in pulse [5,6] and CW [3] EPR. When the modulation frequency exceeds the EPR linewidth, multi-photon transitions result in modulation sidebands in the EPR spectrum. For linewidths larger than the modulation frequency the sidebands are not resolved. It has been shown [3] that both the derivative shape of the lines in standard CW EPR spectra and the distortions due to overmodulation are caused by the unresolved multi-photon transitions. It was also found that the saturation of multi-photon transitions with the different sideband index $k$ may be significantly different. However, an analytical description of multi-photon transitions in conventional CW EPR has been developed only for $\omega_{rf} \gg \omega_1$, where $\omega_1$ is the Rabi frequency in the MW field.

In particular, the failure of this description occurs at the so-called Rabi resonance, when the modulation frequency matches the Rabi frequency, $\omega_1 \sim \omega_{rf}$. This resonance termed 'rotary saturation' has been observed by Redfield in CW NMR [7]. In Redfield's experiment, the dispersion NMR signal arising from a strong incident RF field was observed as a function of the frequency of a weaker audio-frequency field oriented along the static magnetic field. In the NMR experiment with strong and weak transverse RF fields [8], resonance conditions for the weak-field absorption have been studied, when the frequency difference of the two fields was close to the strong-field Rabi frequency. The similar experiment with two transverse microwave fields has been performed in multi-quantum CW EPR [9]. In this experiment an analog of the rotary saturation was realized. The resonance at the Rabi frequency has been observed in pulse NMR, using a phase modulation of the RF field [10], and, recently, in CW NMR using an amplitude-modulated RF field [11]. An optical analog of this phenomenon has been demonstrated at a phase [12,13] and amplitude [14] modulated excitation of two-level atom. In quantum optics, this phenomenon has also been investigated in the time-resolved [15] and spectral CW [16] experiments when two-level atoms were driven by a bichromatic light containing a strong resonant component and a weaker off-resonance component detuned from atomic resonance by the strong-field Rabi frequency. In recent years, the resonance at the Rabi frequency has been studied in time-resolved EPR [17,18] and NMR [19] experiments. Such doubly resonant interaction between the bichromatic (transverse MW and longitudinal RF) field and a two-level spin system (spin qubit) can be described in terms of doubly dressed states and results in an additional Rabi oscillations caused by transitions between these states. These studies are of interest due to their potential applications in quantum information technologies [20,21], for emulation of hybrid spin-mechanical systems [22], for measuring weak RF field [23] as well as in quantum amplifiers and attenuators [24] or in microwave quantum photonics [25].

In this paper we extend the theoretical description of MW and RF multi-photon transitions between the doubly dressed states of spin qubits to the Rabi resonance condition realized in conventional CW EPR spectroscopy with a magnetic field modulation. To our knowledge, the multi-photon transitions under this condition have not been studied in such spectroscopic technique until now. Using the non-secular perturbation theory based on the Bogoliubov averaging method, we obtain the analytical description of absorption EPR spectra and demonstrate the influence of the Rabi resonance for all sidebands in all modulation frequency harmonics. We illustrate the obtained results for experimental conditions corresponding to traditional CW EPR with the 100 kHz modulation frequency for samples with long relaxation times and narrow EPR lines, such as paramagnetic nitrogen centers in diamonds. The significantly different MW saturation for the in-phase and out-of-phase EPR signals can be used to optimize the signal-noise ratio for such samples. We find that the out-of-phase absorption first-harmonic EPR signal is considerably stronger than the in-phase one. Another interesting result is the inversion of two-photon ($k=1$) lines in some spectral intervals of the out-of-phase first-harmonic signal under passage through the Rabi resonance. The manifestation of this effect in some published EPR experiments is discussed. Additionally, we demonstrate the resonance at the Rabi frequency in the time-resolved EPR signals resulting from the total response of the spin system at all modulation frequency harmonics.

## 2. Theory

We consider a spin qubit with the ground $|1\rangle$ and excited $|2\rangle$ states in three fields: a MW one oriented along the $x$ axis of the laboratory frame together with RF and static magnetic fields, both directed along the $z$ axis. The Hamiltonian of the qubit can be written as

$$H = \omega_0 s^z + \omega_1 (s^+ + s^-)\cos\omega_{mw}t + 2\omega_2 s^z \cos\omega_{rf}t, \qquad (1)$$

where $\omega_0 = \gamma B_0$ is the Larmor frequency of the electron spin in the static magnetic field $B_0$, $\gamma$ is the electron gyromagnetic ratio, $2\omega_1 = \gamma B_1$, $2\omega_2 = \gamma B_2$, $B_1$ and $B_2$ are the amplitudes of linearly polarized MW and RF fields, respectively, whereas $s^{\pm,z}$ are components of the spin operator, describing the state of the qubit and satisfying the commutation relations: $[s^+, s^-] = 2s^z$, $[s^z, s^{\pm}] = \pm s^{\pm}$. Since usually $\omega_1/\omega_{mw} \ll 1$, the rotating-wave approximation (RWA) is used for the interaction between the qubit and the MW field.

The dynamics of the qubit is described by the master equation for the density matrix $\rho$

$$i\hbar\frac{\partial\rho}{\partial t} = [H,\rho] + i\Lambda\rho \qquad (2)$$

(in the following we take $\hbar = 1$). The superoperator $\Lambda$ describing decay processes is defined as

$$\Lambda\rho = \frac{\gamma_{21}}{2}D[s^-]\rho + \frac{\gamma_{12}}{2}D[s^+]\rho + \frac{\eta}{2}D[s^z]\rho, \quad (3)$$

where $D[O]\rho = 2O\rho O^+ - O^+O\rho - \rho O^+O$, $\gamma_{21}$ and $\gamma_{12}$ are the rates of the transitions from the excited state $|2\rangle$ of the qubit to its ground state $|1\rangle$ and vice versa, and $\eta$ is the dephasing rate.

We consider two different regimes of spin dynamics which can be realized in CW EPR experiments by proper choice of the modulation frequency and amplitude as well as the MW amplitude. We distinguish weak modulation near the Rabi resonance ($\omega_2 \ll \omega_1 \sim \omega_{rf}$) and strong and fast modulation ($\omega_2 > \omega_1 \ll \omega_{rf}$).

## 2.1. Weak modulation near Rabi resonance

We assume that $\omega_2 \ll \omega_1 \sim \omega_{rf}$ and the MW field is strong, $\omega_1 \gg \gamma_\parallel, \gamma_\perp$, where $\gamma_\parallel = \gamma_{12} + \gamma_{21}$ and $\gamma_\perp = (\gamma_{12} + \gamma_{21} + \eta)/2$ are the rates of energetic and phase relaxation, respectively. The relaxation rates $\gamma_\parallel = 1/T_1$ and $\gamma_\perp = 1/T_2$ represent the longitudinal, $T_1$, and transverse (coherence), $T_2$, relaxation times of the bare qubit in the laboratory frame.

After two canonical transformations $\rho_2 = u_2^+ u_1^+ \rho u_1 u_2$, where $u_1 = \exp(-i\omega_{mw}ts^z)$ and $u_2 = \exp(-i\theta s^y/2)$, Eq. (3) is transformed into $i\partial\rho_2/\partial t = [H_2, \rho_2] + i\Lambda'\rho_2$. Here

$$H_2 = \Omega s^z - \omega_2 \sin\theta \cos(\omega_{rf}t)(s^+ + s^-) + 2\omega_2 \cos\theta \cos(\omega_{rf}t)s^z, \quad (4)$$

$$\Lambda'\rho_2 = \frac{\Gamma_\downarrow}{2}D[s^-]\rho_2 + \frac{\Gamma_\uparrow}{2}D[s^+]\rho_2 + \frac{\Gamma_\varphi}{2}D[s^z]\rho_2, \quad (5)$$

$$\Gamma_\downarrow = \frac{1}{4}(\gamma_{21} + \gamma_{12})(1 + \cos^2\theta) + \frac{1}{2}(\gamma_{21} - \gamma_{12})\cos\theta + \frac{1}{4}\eta\sin^2\theta,$$

$$\Gamma_\uparrow = \frac{1}{4}(\gamma_{21} + \gamma_{12})(1 + \cos^2\theta) - \frac{1}{2}(\gamma_{21} - \gamma_{12})\cos\theta + \frac{1}{4}\eta\sin^2\theta,$$

$$\Gamma_\varphi = \eta\cos^2\theta + (\gamma_{21} + \gamma_{12})\sin^2\theta.$$

Moreover, $\Omega = (\omega_1^2 + \Delta^2)^{1/2}$ is the generalised Rabi frequency in the MW field, $\Delta = \omega_0 - \omega_{mw}$, $\cos\theta = \Delta/\Omega$, and $\sin\theta = \omega_1/\Omega$. New relaxation rates can be introduced: $\Gamma_\parallel = \Gamma_\downarrow + \Gamma_\uparrow = \gamma_\parallel + (\gamma_\perp - \gamma_\parallel)\sin^2\theta$, $\Gamma_\perp = (\Gamma_\downarrow + \Gamma_\uparrow + \Gamma_\varphi)/2 = \gamma_\perp - (1/2)(\gamma_\perp - \gamma_\parallel)\sin^2\theta$. These dressed relaxation rates arise due to taking into account the strong interaction between the qubit and the MW field. Since in the case under consideration $\gamma_\parallel, \gamma_\perp \ll \omega_{rf}, \Omega$, the RWA is used to obtain the operator (5) and the terms containing the products of spin operator pairs $s^\pm$ and $s^z$, $s^+$ and $s^+$, $s^-$ and $s^-$ are neglected. After the unitary transformations, the operator (3) retains its form, but the dissipation and phase relaxation rates (5) are changed. Using the canonical transformation $u_3 = \exp\{-i[\Omega t + (2\omega_2 \cos\theta/\omega_{rf})\sin(\omega_{rf}t)]s^z\}$ we move into the frame rotating with the frequency $\Omega$ and eliminate the diagonal interaction term in Hamiltonian (4). Then for $\rho_3 = u_3^+ \rho_2 u_3$ we obtain: $i\partial\rho_3/\partial t = [H_3, \rho_3] + i\Lambda'\rho_3$, where

$$H_3 = -\omega_2 \sin\theta \cos(\omega_{rf}t)\left\{s^+ \sum J_n(a)\exp\left[i((k+n)\omega_{rf} + \delta_k)t\right] + h.c.\right\}, \quad a = 2\omega_2 \cos\theta/\omega_{rf},$$

$\delta_k = \Omega - k\omega_{rf} \ll \omega_{rf}$ ($k=1,2,3...$), and $J_n$ is the Bessel function of the first kind.

Using the Bogoliubov method (see, e.g. [6, 26]) for averaging over the period $2\pi/\omega_{rf}$, Hamiltonian $H_3$ can be replaced by the effective one up to the second order in a small parameter $\omega_2 \sin\theta/\omega_{rf}$:

$H_3 \to H_3^{eff}(k) = H_3^{(1)}(k) + H_3^{(2)}(k)$, where

$$H_3^{(1)}(k) = \frac{\omega_1 \omega_{rf} k}{2\Delta} J_{-k}(a)(s^+ e^{i\delta_k t} + h.c.), \quad H_3^{(2)}(k) = \Delta_{BS}(k)s^z, \tag{6}$$

$$\Delta_{BS}(k) = \frac{\omega_2^2 \sin^2\theta}{2\omega_{rf}}\left[\sum_{n\neq -k+1}\frac{1}{n+k-1}\left(J_n^2(a) + J_n(a)J_{n-2}(a)\right) + \sum_{n\neq -k-1}\frac{1}{n+k+1}\left(J_n^2(a) + J_n(a)J_{n+2}(a)\right)\right]$$

is the Bloch–Siegert shift. After the transformation $\rho_3 \to \rho_{4,k} = u_4^+ \rho_{3,k} u_4$ with $u_4 = e^{i\delta_k ts^z}$, we obtain $i\partial\rho_{4,k}/\partial t = [H_4, \rho_{4,k}] + i\Lambda'\rho_{4,k}$, where $H_4(k) = \square_k s^z + (w_k/2)(s^+ + s^-)$, $\square_k = \Omega - k\omega_{rf} + \Delta_{BS}(k)$, and $w_k = (\omega_1 \omega_{rf} k/\Delta)J_{-k}(a)$. A steady-state solution of the obtained master equation can be written as $\rho_{4,k}^{st}(t) = 1/2 + \alpha_k s^+ + \alpha_k^* s^- + \beta_k s^z$, where $\alpha_k = \sigma_0(\square_k + i\Gamma_\perp)w_k f_k/2\Gamma_\perp^2$, $\beta_k = \sigma_0(1 + \square_k^2/\Gamma_\perp^2)f_k$. Here we introduce $f_k = (1 + w_k^2/\Gamma_\perp \Gamma_\square + \square_k^2/\Gamma_\perp^2)^{-1}$ and $\sigma_0 = -(\Gamma_\downarrow - \Gamma_\uparrow)/\Gamma_\square$.

In a frame rotating with $\omega_{mw}$ around the $z$ axis of the laboratory frame, the steady-state solution can be written in the following form:

$$\rho_{1,k}^{st}(t) = u_2 u_3 u_4 \rho_{4,k}^{st}(t) u_4^+ u_3^+ u_2^+ = \frac{1}{2}\Big\{1 + \Big[\alpha_k\big((\cos\theta+1)s^+ + (\cos\theta-1)s^- - 2s^z \sin\theta\big)\exp\big(-i(k\omega_{rf} + a\sin\omega_{rf}t)\big) + h.c.\Big] + \beta_k\big((s^+ + s^-)\sin\theta + 2s^z \cos\theta\big)\Big\}. \tag{7}$$

Using Eq. (7), we obtain the EPR absorption signal summing over all $k$-photon resonances in the rotary frame:

$$V^{st} = \frac{1}{2i}\sum_k \left(\langle 1|\rho_{1,k}^{st}|2\rangle - \langle 2|\rho_{1,k}^{st}|1\rangle\right) = \sum_{k,n} J_n(a)\left[\text{Re}(\alpha_k)\sin(k+n)\omega_{rf}t - \text{Im}(\alpha_k)\cos(k+n)\omega_{rf}t\right].$$

(8)

For a measured $j$-th harmonic of this signal only the terms of $V^{st}$ oscillating with the frequency $j\omega_{rf}$ have to be taken into account. In this case the relation $|k+n|=j$ must be fulfilled. Then we obtain

$$V_j^{st} = \sum_k \text{Re}(\alpha_k)\left[J_{j-k}(a) - J_{-j-k}(a)\right]\sin(j\omega_{rf}t) - \sum_k \text{Im}(\alpha_k)\left[J_{j-k}(a) + J_{-j-k}(a)\right]\cos(j\omega_{rf}t). \qquad (9)$$

The in-phase ($A_j^0$) and out-of-phase ($A_j^{\pi/2}$) components of the $j$-th harmonic absorption signal can be written as

$$A_j^0 = -\sigma_0 \sum_k \frac{w_k}{2\Gamma_\perp} \frac{J_{j-k}(a) + J_{-j-k}(a)}{1 + w_k^2/\Gamma_\perp\Gamma_\square + \square_k^2/\Gamma_\perp^2}, \quad A_j^{\pi/2} = \sigma_0 \sum_k \frac{\square_k w_k}{2\Gamma_\perp^2} \frac{J_{j-k}(a) - J_{-j-k}(a)}{1 + w_k^2/\Gamma_\perp\Gamma_\square + \square_k^2/\Gamma_\perp^2}. \qquad (10)$$

In-phase and 90-out-of-phase (or simply out-of-phase) components refer to the phase of the EPR signal relative to that of the magnetic field modulation.

*2.2. Strong-fast modulation*

We assume that $\omega_2 > \omega_1 \ll \omega_{rf}$. In this case we use the canonical transformation of the density matrix $\rho \to \rho_1 = u_1^+ \rho u_1$, where $u_1 = \exp\{-i[\omega_{mw}t + \bar{a}\sin(\omega_{rf}t)]s^z\}$, $\bar{a} = 2\omega_2/\omega_{rf}$. After such transformation, Eq. (2) can be rewritten as: $i\partial\rho_1/\partial t = [H_1, \rho_2] + i\Lambda\rho_1$, where $H_1 = (\omega_1/2)s^+ \exp\left[i\left(\Delta t + \bar{a}\sin(\omega_{rf}t)\right)\right] + h.c.$ Using the Bogoliubov method (see, e.g. [6,26]) for averaging over the period $2\pi/\omega_{rf}$, Hamiltonian $H_1$ can be replaced by the effective one up to the second order in $\omega_1/\omega_{rf}$: $H^{eff}(k) = (\bar{w}_k/2)(e^{-i\bar{\delta}_k t}s^+ + h.c.) + \bar{\Delta}_{BS}(k)s^z$. The corresponding master equation is $i\partial\rho_k^{eff}/\partial t = \left[H^{eff}(k), \rho_k^{eff}\right] + i\Lambda\rho_k^{eff}$, where $\bar{w}_k = \omega_1 J_{-k}(\bar{a})$, $\bar{\Delta}_{BS}(k) = \frac{1}{2}\sum_{n\neq k}\frac{\omega_1^2}{(k-n)\omega_{rf}}J_n^2(\bar{a})$ is the Bloch–Siegert shift, $\bar{\delta}_k = k\omega_{rf} - \Delta$, $k = 0, \pm 1, \pm 2,...$ In the frame rotating with $\omega_{mw}$, a steady-state solution of the master equation can be expressed as

$$\rho_{rot,k}^{st}(t) = \frac{1}{2} + \left\{\bar{\alpha}_k s^+ \exp\left[-i(k\omega_{rf}t + \bar{a}\sin(\omega_{rf}t))\right] + h.c.\right\} + \bar{\beta}_k s^z, \qquad (11)$$

where $\square_k = \Delta - k\omega_{rf} + \bar{\Delta}_{BS}(k)$, $\bar{\alpha}_k = \bar{\sigma}_0(\square_k + i\gamma_\perp)\bar{w}_k\bar{f}_k/2\gamma_\perp^2$, $\bar{\beta}_k = \bar{\sigma}_0\left(1 + \square_k^2/\gamma_\perp^2\right)\bar{f}_k$, $\bar{\sigma}_0 = -(\gamma_{21} - \gamma_{12})/\gamma_\square$, and $\bar{f}_k = (1 + \bar{w}_k^2/\gamma_\perp\gamma_\square + \square_k^2/\gamma_\perp^2)^{-1}$. Using Eq. (7), we obtain the $j$-th harmonic absorption signal:

$$\bar{V}_j^{st} = \bar{A}_j^{\pi/2}\sin(j\omega_{rf}t) + \bar{A}_j^0\cos(j\omega_{rf}t), \qquad (12)$$

where $\bar{A}_j^0 = -\bar{\sigma}_0 \sum_k \frac{\bar{w}_k}{2\gamma_\perp} \frac{J_{j-k}(\bar{a}) + J_{-j-k}(\bar{a})}{1 + \bar{w}_k^2/\gamma_\perp\gamma_\square + \square_k^2/\gamma_\perp^2}$, $\bar{A}_j^{\pi/2} = \bar{\sigma}_0 \sum_k \frac{\square_k \bar{w}_k}{2\gamma_\perp^2} \frac{J_{j-k}(\bar{a}) - J_{-j-k}(\bar{a})}{1 + \bar{w}_k^2/\gamma_\perp\gamma_\square + \square_k^2/\gamma_\perp^2}$.

Apart from the neglected Bloch–Siegert shift, Eq. (12) is identical to expression (59) obtained in [3].

## 3. Applications to CW EPR with 100 kHz modulation

In this section, we provide examples on the application of the obtained analytical expressions. We use parameters corresponding to traditional CW EPR experiments with the 100 kHz modulation frequency for samples with long relaxation times and narrow EPR lines. At room temperature such relaxation times and linewidths are observed for some defects in solids, especially for isolated neutral substitutional nitrogen (P1 center) and the negatively charged nitrogen vacancy center in diamonds. The latter centers are particularly promising quantum systems (solid-state qubits) with wide applications in physics, quantum information processing and biology [27]. The coherence time $T_2$ of these paramagnetic centers depends on their concentration and can be changed from a few microseconds to a few milliseconds in ultrapure diamonds [27-30]. In the latter case, $T_2$ becomes comparable with the spin-lattice relaxation time $T_1$ [28,30]. The parameters of these centers allow us to resolve the modulation sidebands originated from the multi-photon transitions and to demonstrate the transformation of these sidebands under the Rabi resonance in traditional CW EPR experiments with the 100 kHz modulation frequency.

### 3.1. First-harmonic in-phase spectrum

Figures 1 and 2 depict the first-harmonic in-phase absorption EPR spectra. At first we present the multi-photon EPR signals that are characteristic far from the Rabi resonance ( $\omega_2 > \omega_1 \ll \omega_{rf}$ ). Fig. 1 shows the first-harmonic in-phase absorption EPR spectra calculated from Eq. (12) for different values of the Rabi frequency $\omega_1$. These signals were calculated for strong-fast modulation. In order to observe resolved multi-photon transitions, the modulation frequency exceeds the EPR linewidth. In this case the obtained EPR spectra consist of sidebands with absorptive lineshape. The sidebands are found at $\Delta = k\omega_{rf}$. The single-photon transition ( $k = 0$ ) gives no contribution to this signal. We observe the broadening and saturation of the sidebands with increasing $\omega_1$. The effective saturation and the saturation broadening of sidebands depends on the sideband index $k$. When $\omega_1$ increases, the sidebands with $k = \pm 1$ broaden and are saturated stronger than the sidebands with higher $k$.

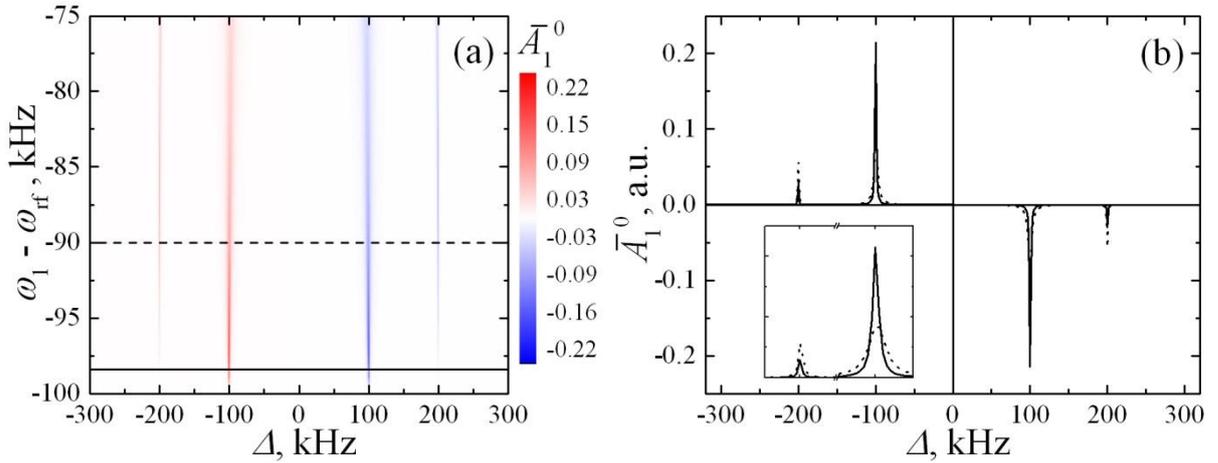

Fig. 1. (a) The first-harmonic in-phase absorption EPR signal calculated from Eq. (12) as a function of $\omega_1$ for $\omega_2 > \omega_1 \ll \omega_{rf}$. $\omega_{rf} = 2\pi\ 100$ kHz, $\omega_2 = 2\pi\ 28$ kHz, $\gamma_\parallel = \gamma_\perp = 0.5$ kHz. (b) Two cuts of (a) at $\omega_1 = 2\pi\ 3.0$ kHz (solid line) and $\omega_1 = 2\pi\ 10$ kHz (dashed line). The same lines show positions of these spectra in (a). The inset shows the sidebands in more detail.

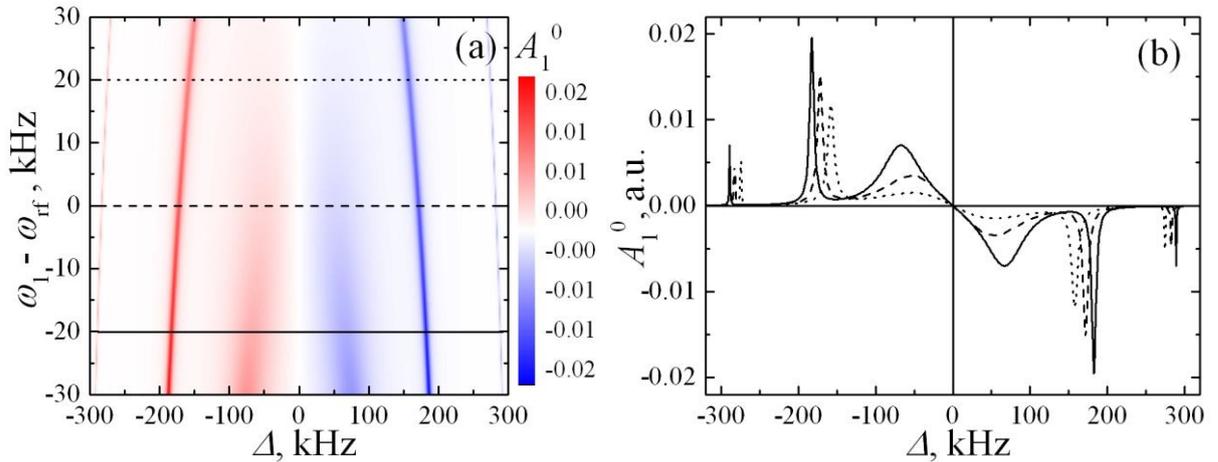

Fig. 2. (a) The first-harmonic in-phase absorption EPR signal calculated from Eq. (10) as a function of $\omega_1$ for $\omega_2 \ll \omega_1 \sim \omega_{rf}$. (b) Three cuts of (a) at $\omega_1 = 2\pi\ 80$ kHz (solid line), $\omega_1 = 2\pi\ 100$ kHz (dashed line) and $\omega_1 = 2\pi\ 120$ kHz (dotted line). The same lines show positions of these spectra in (a).

The first-harmonic in-phase absorption EPR signals calculated from Eq. (10) near the Rabi resonance ($\omega_2 \ll \omega_1 \sim \omega_{rf}$) are illustrated in Fig. 2. In contrast to the previous case ($\omega_2 > \omega_1 \ll \omega_{rf}$) when each multi-photon transition can be considered separately, the different-order transitions begin to overlap due to the strong saturation broadening of sidebands (Fig. 2b). Moreover, the sidebands shift from $\Delta = k\omega_{rf}$ with increasing $\omega_1$ and they occur at $|\Delta| < k\omega_{rf}$.

These shifts are smaller for the sidebands with higher $k$ than for the sidebands with $k=1$. In addition, the sidebands with $k=1$ become weak due to its stronger saturation and they are very weak at $\omega_1 > \omega_{rf}$. Due to the Bloch–Siegert effect the sidebands are additionally shifted, and the resonance occurs at $\omega_1$ a little lower than $\omega_{rf}$. In our case the additional shift is of about 2 kHz for $k=1$.

Fig. 3 shows multi-photon transitions in the rotary frame and positions of sidebands presented in Fig 2. Vertical arrows show absorbed ($\Delta > 0$) or emitted ($\Delta < 0$) RF photons for $\omega_1 < \omega_{rf}$ and $\omega_1 > \omega_{rf}$. For example, the two-photon process in the laboratory frame when one MW photon is absorbed simultaneously with absorption of one RF photon corresponds to the absorption of one RF photon in the rotary frame. At $\omega_1 < \omega_{rf}$ the sideband positions are found from the multi-photon resonance condition $\square_k = \Omega - k\omega_{rf} + \Delta_{BS}(k) = 0$, where $\Omega = \left(\omega_1^2 + \Delta^2\right)^{1/2}$. The one-photon process in the rotary frame ($k=1$) can be realized even when $\omega_1 > \omega_{rf}$ (Fig. 3b) because the frequency of the RF photon is comparable with the width of the quantum transition. This width is proportional to $\left(\Gamma_\perp^2 + w_1^2 \Gamma_\perp / \Gamma_\parallel\right)^{1/2}$ and is shown in Fig. 3b by vertical grey bands. One can see that for $k \geq 2$ these widths are significantly smaller and does not influence the realization of the multi-photon resonance condition. So, at the non-resonant RF modulation ($\omega_2 > \omega_1 \ll \omega_{rf}$), the sidebands shift from $\Delta = k\omega_{rf}$ only due to the Bloch–Siegert effect and these shifts are small (< 1 kHz), as it is illustrated in Fig. 1. At the resonant modulation ($\omega_2 \ll \omega_1 \sim \omega_{rf}$), the strong MW field modifies the multi-photon resonance condition and shifts significantly the sidebands from $\Delta = k\omega_{rf}$.

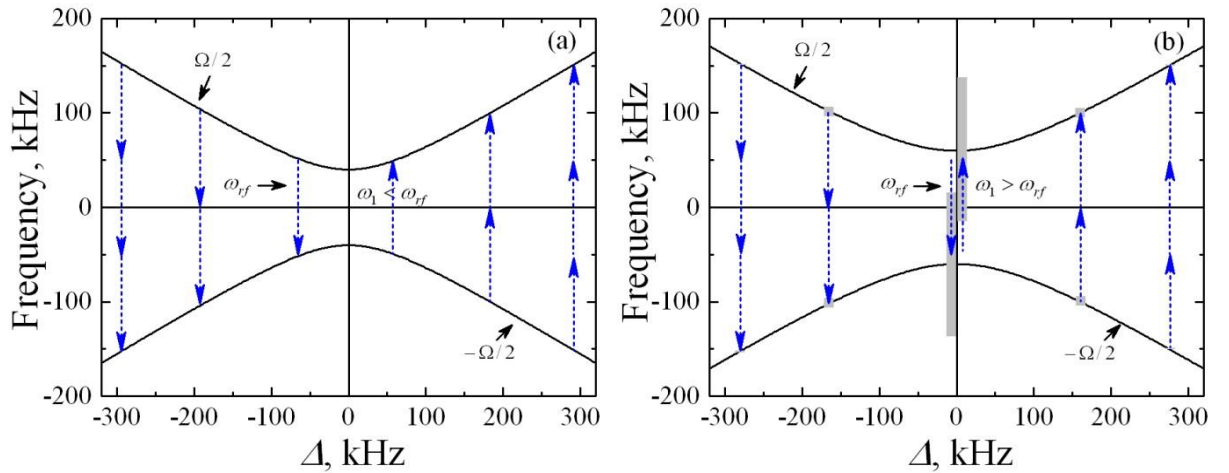

Fig. 3. Multi-photon transitions in the rotary frame for the spectra presented in Fig. 2 with $\omega_1 = 2\pi\, 80$ kHz (a) and $\omega_1 = 2\pi\, 120$ kHz (b) for $\omega_2 \ll \omega_1 \sim \omega_{rf}$. Vertical arrows show absorbed or emitted RF photons. Vertical grey bands present widths of quantum transitions.

### 3.2. First-harmonic out-of-phase spectrum

For $\omega_2 > \omega_1 \ll \omega_{rf}$, the first-harmonic $90^0$-out-of-phase spectrum consists of a central line at $\Delta = 0$ and sidebands at $\Delta = k\omega_{rf}$ (Fig. 4). All components have dispersive shape. The spectrum is anti-symmetric with respect to the central line. The effective saturation and the saturation broadening of components depends on the sideband index $k$. The saturation of the central line caused by the single–photon transition is stronger than the saturation of sidebands resulting from the multi-photon transitions (Fig. 4b).

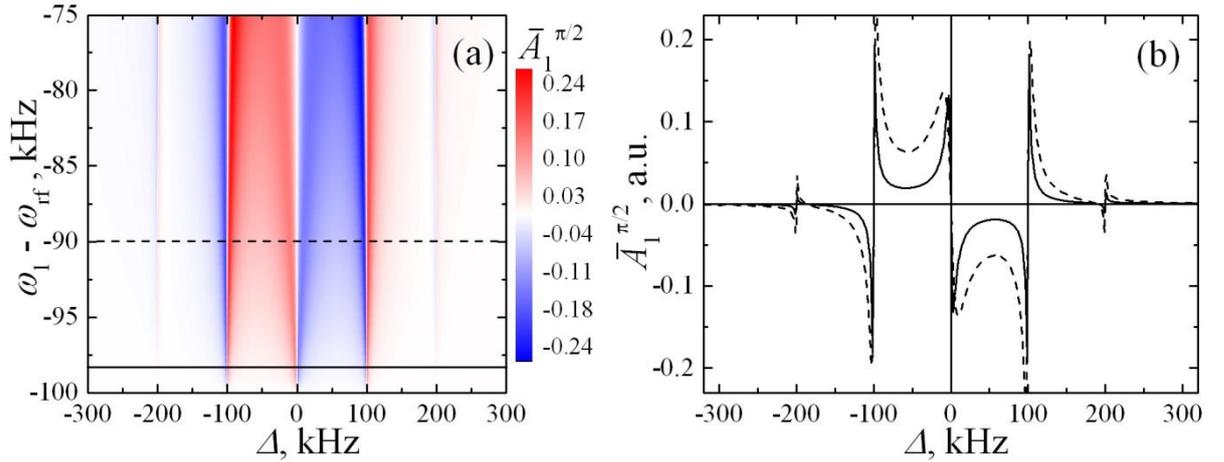

Fig. 4. (a) The first-harmonic out-of-phase absorption EPR signal calculated from Eq. (12) as a function of $\omega_1$ for $\omega_2 > \omega_1 \ll \omega_{rf}$. (b) Two cuts of (a). Other parameters and denotations are the same as in Fig. 1.

At $\omega_2 \ll \omega_1 \sim \omega_{rf}$, noticeable changes of the out-of-phase spectrum are observed (Fig. 5). Unlike the out-of-phase spectrum at $\omega_2 > \omega_1 \ll \omega_{rf}$, there is no signal of the single-photon transition ($k = 0$) because $w_{k=0} = 0$. The different-order sidebands begin to overlap due to the strong broadening. All sidebands are shifted from $|\Delta| = k\omega_{rf}$ and occur at $|\Delta| < k\omega_{rf}$. It is follows from Eq. (10) that the sideband positions (zero values) shown in white in Fig. 5a are given by the following resonance condition in the rotating frame: $\square_k = \Omega - k\omega_{rf} + \Delta_{BS}(k) = 0$. In addition, the shape of each sideband is changed appreciably when the detuning $\omega_1 - \omega_{rf}$ is varied. The strongest changes occur for the sidebands formed by two-photon transitions, where one MW photon is absorbed simultaneously with absorption or emission of one RF photon (Fig. 5b). Because at $\Delta = \pm\left[\left(\omega_{rf} - \Delta_{BS}(1)\right)^2 - \omega_1^2\right]^{1/2}$ $\square_1$ equals to zero and changes its sign when the detuning $\Delta$ is varied, the lineshape of these sidebands is transformed from a nearly dispersive (at $\omega_1 < \omega_{rf}$) to absorptive (at $\omega_1 > \omega_{rf}$) shape. Since at $\omega_1 > \omega_{rf}$ $\square_1$ does not changes its sign, the amplitude of the two-photon line ($k=1$) is positive for $\Delta > 0$ and is negative for $\Delta < 0$. The shape transformation of these sidebands demonstrates that their phase varies in some spectral intervals as $\omega_1$ increases and passes through the Rabi resonance. For the sidebands with higher $k$ these changes are smaller and they keep nearly dispersive lineshape. Note that the out-phase signals with $k=1$ are stronger (about 20 times) than those detected in-phase. Therefore, the out-phase signal is more appropriate to identify the resonance at the Rabi frequency.

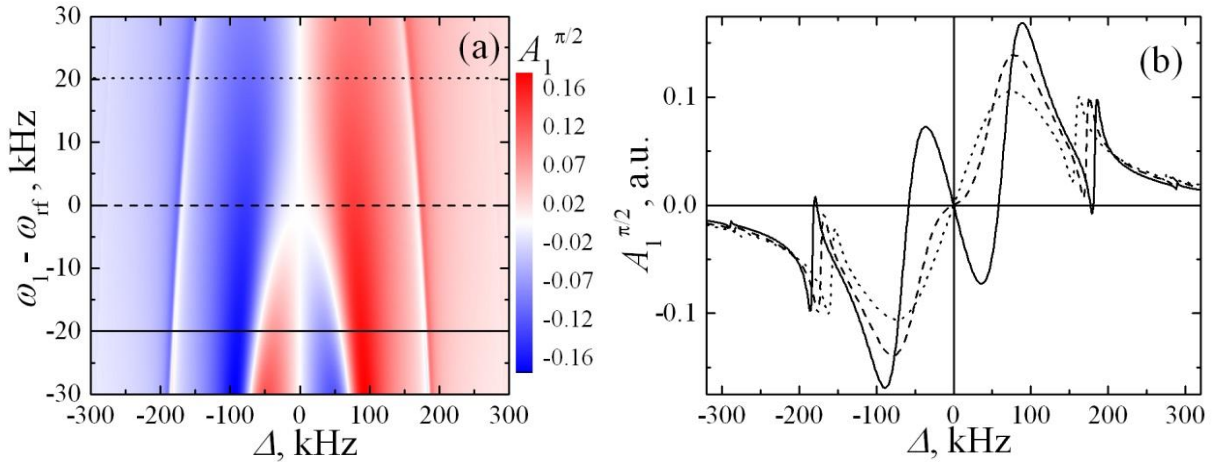

Fig. 5. (a) The first-harmonic out-of-phase absorption EPR signal calculated from Eq. (10) as a function of $\omega_1$ for $\omega_2 \ll \omega_1 \sim \omega_{rf}$. (b) Three cuts of (a). Other parameters and denotations are the same as in Fig. 2.

Fig. 5 demonstrates the changes of the signs of two-photon lines ($k=1$) in some spectral intervals of the out-of-phase EPR signal when the detuning $\omega_1 - \omega_{rf}$ is varied. The change of the sign of the absorption EPR signal (line inversion) indicate the transition in these intervals from absorption to stimulated emission for $\Delta < 0$ or vice versa for $\Delta > 0$.

At $\omega_1 > \omega_{rf}$ each of the sidebands with $k=1$ is transformed to absorptive shape. These sidebands are anti-symmetric with respect to $\Delta = 0$ and have opposite phases. Due to broadening, together these two sidebands look as one broad dispersive line (dotted line in Fig. 5b) or the derivative line commonly observed in CW EPR. We can compare this EPR signal with the signal formed at $\omega_2 > \omega_1 \ll \omega_{rf}$ in the same spectral interval, $|\Delta| < 100$ kHz. In the last case there is the narrow dispersive line caused by the single–photon transition with $k = 0$ (the central line at $\Delta = 0$ in Fig. 4b). The broad dispersive line observed at $\omega_1 > \omega_{rf}$, $\omega_2 \ll \omega_1$ and the narrow dispersive line observed at $\omega_2 > \omega_1 \ll \omega_{rf}$ have opposite phases. Consequently, an increase of $\omega_1$ from the low ($\omega_2 > \omega_1 \ll \omega_{rf}$) to high ($\omega_1 > \omega_{rf}$, $\omega_2 \ll \omega_1$) values also causes the inversion of the out-of-phase absorption EPR signals. This inversion results from the photon transitions with the different sideband index $k$, $k = 0$ and $k = 1$ for the low to high values of $\omega_1$, respectively.

### 3.3. Second-harmonic spectrum

Fig. 6 depicts the in-phase and out-of-phase second-harmonic spectra near the Rabi resonance. These spectra are symmetric with respect to the central line. Like the first-harmonic spectra, the lineshapes of sidebands and their positions are changed as $\omega_1$ passes through the resonance condition. However, the changes are smaller in comparison with the first-harmonic spectra.

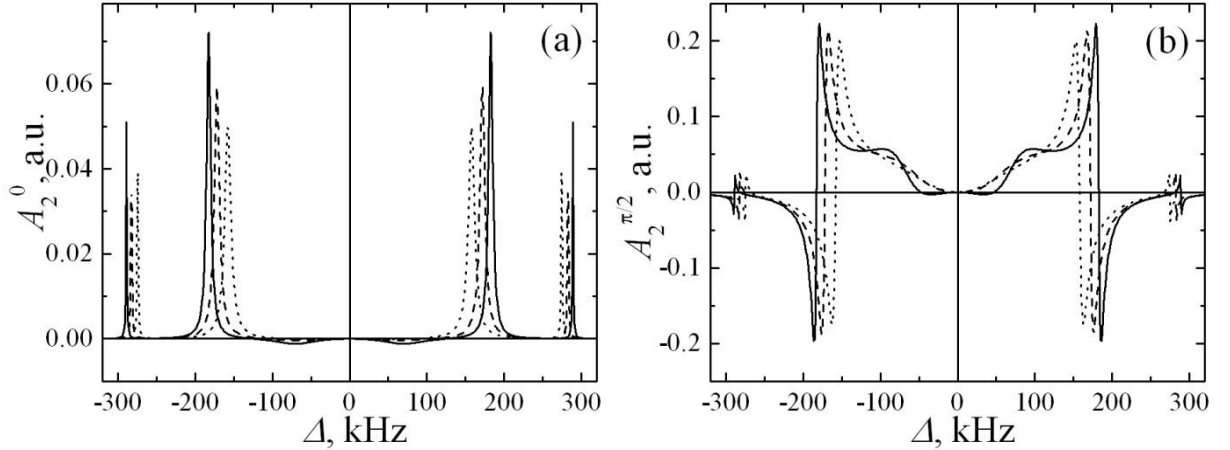

Fig. 6. The in-phase (a) and out-of-phase (b) second-harmonic spectra near the Rabi resonance. Other parameters and denotations are the same as in Fig. 2.

### 3.4. Time dependence of CW absorption signals under Rabi resonance

The time-resolved absorption EPR signals (waveforms) permit an analysis of the total response of the spin system near the Rabi resonance (Fig. 7). Each waveform reflects the steady state behavior of the spin system and was calculated from Eq. (8) at the fixed values of the detuning $\Delta = \omega_0 - \omega_{mw}$. This signal results from the spin response at all harmonics of the modulation frequency and can be observed prior to phase-sensitive detection. The time behavior of the EPR signals demonstrates the contribution from higher-order harmonics. The contribution depends on both the detuning $\omega_1 - \omega_{rf}$ and the detuning $\Delta = \omega_0 - \omega_{mw}$. It is most obviously revealed when the first-harmonic signal becomes weak due to the fulfillment of the resonance condition $\square_k = \Omega - k\omega_{rf} + \Delta_{BS}(k) = 0$ for the first harmonic. At $\Delta = -10$ kHz and $\Delta = -25$ kHz the increase of $\omega_1$ from 90 kHz to 110 kHz is quite enough to pass through the resonance given by the condition $\square_1 = 0$. The signal undergoes a $180^0$ change of the phase under the passing. At $\Delta = -40$ kHz, the increase of $\omega_1$ from 90 kHz to 110 kHz does not fulfil the condition $\square_1 = 0$. At these parameters there is no $180^0$ change of the phase of the presented signals.

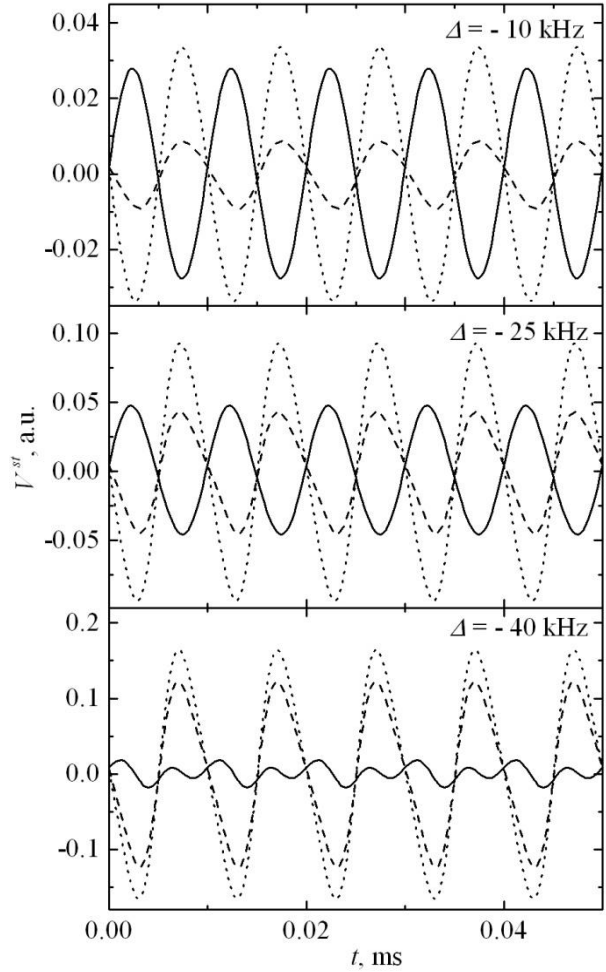

Fig. 7. Time evolution of the absorption signal calculated from Eq. (8) for three values of detuning $\Delta$ at the magnetic field modulation with $\omega_{rf} = 2\pi$ 100 kHz, $\omega_2 = 2\pi$ 28 kHz, $\gamma_\parallel = \gamma_\perp$ = 0.5 kHz, $\omega_1 = 2\pi$ 90 kHz (solid line), $\omega_1 = 2\pi$ 100 kHz (dashed line) and $\omega_1 = 2\pi$ 110 kHz (dotted line).

Fig. 8 demonstrates that the contribution from higher-order harmonics increases as $\omega_1$ and $\Delta$ increases. The signals formed by the higher-order harmonics exceed the first-harmonic signal only at $\Omega_1 = 0$ and $\Delta = -40$ kHz. At smaller $\Delta$, the double saturation in the MW and RF fields is stronger for the first-harmonic signal than for higher-harmonic signals. Note that the $180^0$ change of the phase of the first-harmonic signal for $\Delta = -10$ kHz and $\Delta = -25$ kHz indicates the transition from absorption to stimulated emission. The stimulated emission reduces the strength of the EPR signal. This reducing results in the dip in the dependence of strength of the signals on the Rabi frequency $\omega_1$ (Fig. 8). The position of the dip corresponds to the resonance condition $\Omega_1 = 0$ for the first harmonic.

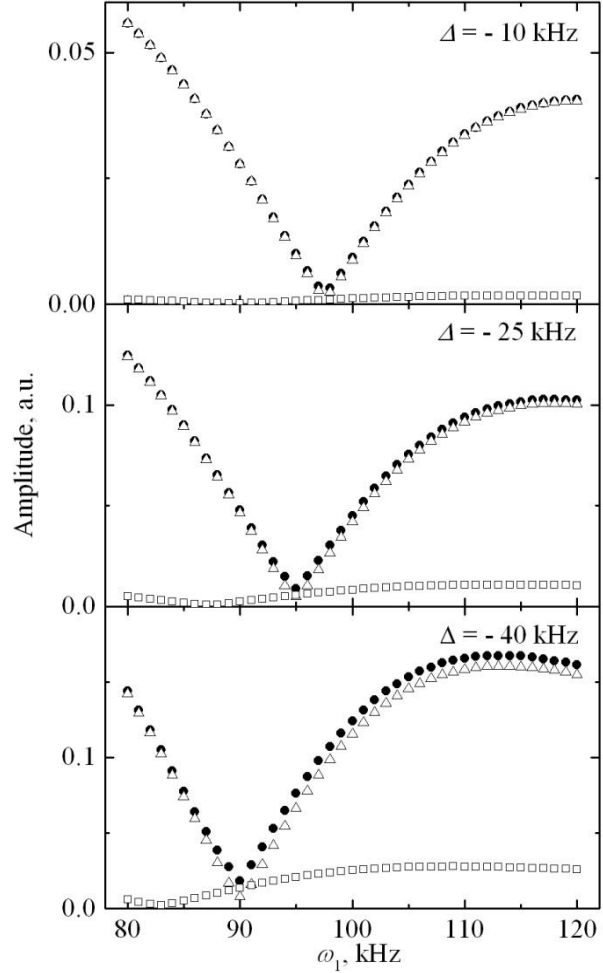

Fig. 8. Strength of the signals as a function of $\omega_1$ for three values of $\Delta$ which are the same as in (a). The circles, triangles and squares are extracted from the total response at all harmonics, from the response at the first harmonic and from the response at the higher-order harmonics, respectively.

*3.5. Secondary Rabi resonances*

As it is demonstrated above, the strong changes in the EPR signal are predicted when the modulation frequency matches the Rabi frequency, $\omega_1 \approx \omega_{rf}$. Besides this primary Rabi resonance, the secondary Rabi resonances are predicted by our theory when the Rabi frequency is equal to $2\omega_{rf}, 3\omega_{rf}$, etc. Fig. 9 depicts the first-harmonic out-of-phase absorption EPR signal calculated from Eq. (10) at $\omega_1 \approx 2\omega_{rf}$. At $\omega_1 \approx 2\omega_{rf}$ two-photon processes (described by the term with $k=1$ in Eq. (10) for $A_1^{\pi/2}$) give the main contribution to the absorption signal. The two-photon signal does not change its sign because $\square_1$ in the numerator of Eq. (10) saves its sign when the detuning $\Delta$ is varied. The three-photon ($k=2$) transitions form the dispersive-like out-of-phase signal due to the existence of the $\Delta$ value for which $\square_2$ changes its sign. However, due to the domination of two-photon processes, the three-photon signals results in a jump of the summarized signal at the $\Delta$ values determined by the condition $\square_2 = 0$. The four-photon signals are visible at $\Delta$ of about $\pm 240$ kHz. The three-photon ($k=2$) and four-photon ($k=3$) signals are too weak to change the sign of the summarized absorption signal. Consequently, unlike the Rabi resonance at $\omega_1 \approx \omega_{rf}$ presented in Fig. 5, there is no line inversion under the passing through the secondary resonance. The dip in the dependence of strength of the signals on the Rabi frequency $\omega_1$ is caused by the secondary Rabi resonance. This dip is smaller in comparison with the dip presented in Fig. 5 for $\omega_1 \approx \omega_{rf}$. Its position corresponds to the resonance condition $\square_2 = 0$ at $\omega_1 \approx 2\omega_{rf}$ for the first harmonic.

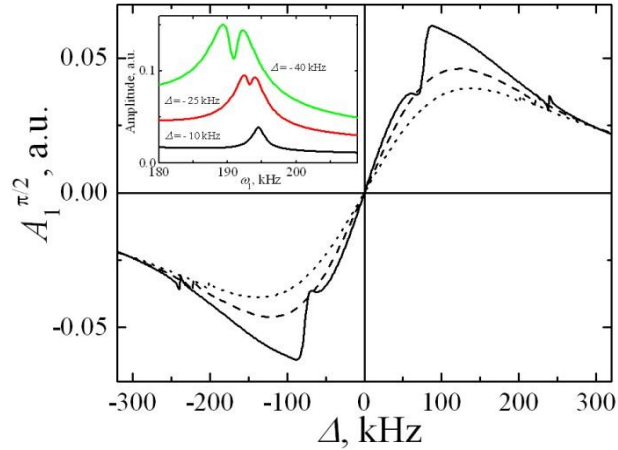

Fig. 9. The first-harmonic out-of-phase absorption EPR signal calculated from Eq. (10) at $\omega_1 \sim 2\omega_{rf}$. $\omega_{rf} = 2\pi\ 100$ kHz, $\omega_1 = 2\pi\ 180$ kHz (solid line), $\omega_1 = 2\pi\ 200$ kHz (dashed line) and $\omega_1 = 2\pi\ 220$ kHz (dotted line). Inset shows strength of the signal as a function of $\omega_1$ for three values of $\Delta$.

## 4. Discussion

The obtained results allow us to disclose the Rabi resonance and multi-photon transitions in traditional CW EPR experiments with a magnetic-field modulation. To our knowledge, this resonance has not been identified in such experiments until now, though the condition of this resonance could be fulfilled at the microwave saturation of the EPR absorption signals for samples with long electron spin relaxation times. Note that in accordance with Eq. (10), there is no the single–photon transition with $k=0$ in the in-phase and out-of-phase absorption EPR signal because $w_{k=0}=0$. Consequently, the resonance at the Rabi frequency is only a multi–photon phenomenon.

Spectra with the well-resolved modulation sideband structure are rare in the EPR literature. The features of sidebands at the Rabi resonance conditions have not been investigated in detail so far. However, the inversion of the out-of-phase absorption EPR signal at increasing of the MW power has been observed for the P1 centers in high-quality diamonds at room temperature using the CW EPR with the 100 kHz modulation [31]. The observed inversion of the EPR signal as well as the strong difference between the in- and out-of-phase signals has not been explained until now. Our results shown in Figs. 4 and 5 were obtained with the parameters close to the experimental conditions in [31] and demonstrate the main features of the observed inversion and the strong difference between the in- and out-of-phase signals. Some differences in the line widths of the calculated and observed signals may result from neglecting the inhomogeneous broadening. The estimated concentration of the P1 centers in the investigated sample was about 4 × $10^{20}$ $m^{-3}$ [31]. The time-resolved EPR measurements have shown that spin-spin relaxation time $T_2$ at such concentration of the P1 centers is longer than one millisecond achieving the value of $T_1$ [28,30]. Therefore, we assumed in the presented illustrations that $T_2=T_1$. Even at room temperature these relaxation times are so long that extremely low microwave powers, modulation amplitudes, and modulation frequencies must be used to record unsaturated and non-distorted CW EPR spectra. Really, it is not possible to obtain such spectra for these diamonds using the 100-kHz modulation. In this case, because the sidebands with higher $k$ broaden and are saturated weaker than the sidebands with $k=1$, the sidebands with higher $k$ can be used to enhance the signal-noise ratio and to detect the unsaturated linewidth. A modulation frequency of 100 kHz limits the observation of resolved sidebands for samples with linewidths larger than 3.5 µT. Modulation frequencies higher than 100 kHz need to resolve multi-photon transitions for such samples.

The predicted inversion of two-photon lines in some spectral intervals of the out-of-phase absorption EPR signal under passage through the Rabi resonance can be a signature of a steady-state population inversion of the doubly dressed–state levels [32]. This inversion can be used for amplification (lasing) without population inversion in the bare-state basis as it was shown in [32] for strongly limited detunings $\Delta \ll \omega_1$. We describe the EPR signal without this limitation.

We present here the analytical description of the Rabi resonance at weak-resonant modulation ($\omega_2 \ll \omega_1 \sim \omega_{rf}$). This resonance can also be realized in the regime of strong-resonant modulation ($\omega_2 \gg \omega_1 \sim \omega_{rf}$), but it is difficult to describe this regime analytically. In our opinion, the resonance at the Rabi frequency in such regime has been experimentally observed in [33]. However, this resonance has not been identified in this paper, though it is obviously revealed as a dip at $\omega_1 \approx \omega_{rf}$ in the dependence of the amplitude of the 100-kHz EPR signal on the MW amplitude $B_1$ (Fig. 5 in [33]). Moreover, the observed strong phase change of the EPR signal at $\omega_1 \approx \omega_{rf}$ (see Fig. 6 in [33]) is other manifestation of the resonance under discussion. In the regime of strong-resonant modulation, the inversion of the in-phase EPR signal of the P1 centers in diamond has also been observed at increasing the MW power [34].

## 5. Conclusion

Using the Bogoliubov averaging method, we have described MW and RF multi-photon transitions excited between the doubly dressed states of spin qubits in CW EPR spectroscopy with a magnetic field modulation for the two cases: (i) weak modulation near the Rabi resonance, $\omega_2 \ll \omega_1 \sim \omega_{rf}$, and (ii) strong-fast modulation, $\omega_2 > \omega_1 \ll \omega_{rf}$. The analytical description of the response of spin system to the CW bichromatic excitation has been obtained for all sidebands, resulting from the multi-photon transitions, in all modulation frequency harmonics. The manifestations of the primary and secondary Rabi resonance have been demonstrated in the spectral response of the spin system observed with phase-sensitive detection at one harmonic as well as in the time-resolved steady-state response formed by all harmonics of the modulation frequency. We have found that the signature of this resonance is the frequency shifts of sidebands. The inversion of two-photon lines in some spectral intervals of the out-of-phase first-harmonic signal was predicted under passage through the Rabi resonance. The dependence of the EPR signal on the phase of the RF modulation and the different saturation of multi-photon transitions with the different sideband index $k$ can be used to optimize the signal-noise ratio for samples with long spin relaxation times and extremely narrow EPR lines. Our results provide a theoretical framework for future developments in multi-photon CW EPR spectroscopy. Their application is not restricted to single electron and nuclear spin qubits or their ensemble. The predicted effects can be realized with a two-level quantum dot or a single superconducting qubit. The found features of multi-photon transitions could also be useful in new applications of non-linear properties of electron spin systems for emulation of hybrid spin-mechanical systems and in quantum amplifiers and attenuators.